\documentclass[preprint, times, twocolumn]{aastex631}
\usepackage{listings}
\usepackage{hyperref}
\usepackage{amssymb}

\newcommand{\wc}[1]{\texcount -v3 -merge -incbib -dir -sub=none -utf8 -sum rnaas.tex}

\shorttitle{iCompare}
\shortauthors{Chernyavskaya et al.}
\graphicspath{{./}{figures/}}

\begin{document}
\title{\texttt{iCompare}: A Package for Automated Comparison of Solar System Integrators \footnote{Released on November 8, 2021; \href{https://zenodo.org/badge/latestdoi/283350311}{doi:10.5281/zenodo.5655805}}}

\correspondingauthor{Maria Chernyavskaya}
\email{maria8ch@uw.edu} 

\author[0000-0002-6292-9056]{Maria Chernyavskaya}
\author[0000-0003-1996-9252]{Mario Juri\'{c}}
\author[0000-0001-5820-3925]{Joachim Moeyens}
\author[0000-0002-1398-6302]{Siegfried Eggl}
\author[0000-0001-5916-0031]{Lynne Jones}

\affiliation{DiRAC Institute and the Department of Astronomy, University of Washington, Seattle, USA}


\begin{abstract}
We present a tool for the comparison and validation of the integration packages suitable for Solar System dynamics. \texttt{iCompare}, written in Python, compares the ephemeris prediction accuracy of a suite of commonly-used integration packages (JPL/HORIZONS, OpenOrb, OrbFit at present). It integrates a set of test particles with orbits picked to explore both usual and unusual regions in Solar System phase space and compares the computed to reference ephemerides. The results are visualized in an intuitive dashboard. This allows for the assessment of integrator suitability as a function of population, as well as monitoring their performance from version to version (a capability needed for the Rubin Observatory's software pipeline construction efforts). We provide the code on GitHub with a readily runnable version in Binder (\url{https://github.com/dirac-institute/iCompare}).
\end{abstract}



\section{Introduction} \label{sec:intro}

Accurate computation of positions and velocities of Solar System objects (ephemerides) is a challenging problem solved by special-purpose numerical integration software packages (integrators). A number of these are in use by the dynamics community: JPL/HORIZONS\footnote{\href{https://ssd.jpl.nasa.gov/?horizons}{https://ssd.jpl.nasa.gov/?horizons}}, OpenOrb\footnote{\href{http://openorb.sourceforge.net/}{http://openorb.sourceforge.net/}} \citep{openorb}, and OrbFit\footnote{\href{http://adams.dm.unipi.it/orbfit/}{http://adams.dm.unipi.it/orbfit/}} to name a few. JPL/HORIZONS is generally regarded as the ``gold standard'' for high-precision short-term computation,
while others may excel for different purposes (e.g., longer-term integration). However, we have found no easy way to compare their performance and select the best tool for any given job. Ideally, one would like a tool that can automatically compute and compare the ephemerides produced by a suite of integrators, and visually present the results. For example, the Rubin Observatory’s Data Management system \citep{2017Juric} for the Legacy Survey of Space and Time (LSST) \citep{LSST-Collab} is required to accurately predict the positions of known moving objects to enable association and monitoring thereof. 

In this note we present \texttt{iCompare}, a software package for comparison and validation of Solar System integrators. Written in Python, it compares the ephemerides prediction accuracies of a suite of commonly-used integrators, and displays the results in a color-coded table.

\section{About \texttt{iCompare}} \label{sec:about}

\subsection{Overview and Architecture}
\texttt{iCompare} is a pure-Python package that automatically runs a suite of orbit integrators with identical inputs and comparable dynamical models; compares their outputs and produces a color-coded table representing the closeness of predicted ephemerides to a reference integrator (Figure~\ref{fig: table1}).

The package is built to be modular and extensible. The main driver and the output visualization functions are all in a single file (\lstinline{icompare/__init__.py}, at present). Code for each supported integrator is implemented in its own file (e.g. \lstinline{icompare/openorb.py}), with the requirement that such modules must export a function with the signature \lstinline{def get_ephem(elements, start, stop, obscode)} where \lstinline{elements} are the orbital elements in cometary format, \lstinline{start} and \lstinline{stop} are the start and stop times of integration, and \lstinline{obscode} is the MPC Observatory Code. This architecture makes it easy to enhance \texttt{iCompare} with support for new integrators\footnote{We welcome pull requests to add new integrators to the comparison suite.}.

\subsection{Supported Integrators and their Configuration} \label{subsec:config}
\texttt{iCompare} supports three integrators at present: the JPL/HORIZONS (the reference integrator), OpenOrb, and OrbFit. These support different options for tuning the dynamical model and integration method. All have been configured to use the DE430 ephemerides, except for JPL/HORIZONS which has recently switched to DE441\footnote{Strictly speaking, the integrators would all use the same dynamical model, or at least the state vectors derived for their particular model. This is not what occurs in common every-day use.}. The details are as follows:

\begin{itemize}
    \item {\bf JPL/HORIZONS} is a web service that offers few user-tunable options. We use it in its default configuration with \lstinline{astroquery.jplhorizons} \citep{astroquery}.
    \item {\bf OpenOrb} uses the default dynamical model of OpenOrb's python bindings, \lstinline{pyoorb}, which is the typical of the way most users would run it. This includes the DE430 ephemerides, N-body integration, a 5-day integration step, and no asteroid perturbers.
    \item {\bf OrbFit} exposes the ability to adjust the dynamical model through numerous settings. Per advice by the OrbFit team (F.~Spoto, priv. comm.), we've used the settings listed in the \lstinline{icompare/orbfit.py} file\footnote{\url{https://github.com/dirac-institute/iCompare/blob/a823cbfeb86bedd40bf4480382e17974f5066827/icompare/orbfit.py##L78}}.
    We caution the reader that these choices may change as the software evolves; the most recent settings can always be found in \lstinline{icompare/orbfit.py}.
\end{itemize}


\subsection{Reference Integrator} \label{subsec: ref}

At present, the reference integrator is fixed to JPL/HORIZONS; a future version will allow it to be user-selectable. All computed ephemerides are compared to JPL/HORIZONS outputs, and the comparisons are displayed in the visual dashboard.


\subsection{Ephemerides comparison} \label{subsec:ephem}
In the default configuration, \texttt{iCompare} computes the predicted positions of given objects over a 10yr timespan, January 1, 2010 -- January 1, 2020, with step size of one day. The default obscode is \lstinline{I11}, the standard stand-in for the Rubin Observatory, Gemini South Telescope.

Once the positions of the propagated objects have been computed, we compute the great circle distances between the predictions of each integrator and the reference integrator \footnote{See Section \ref{subsec: ref}}. These results are passed to \texttt{icompare.table\_stats()}, which computes a median, mean and the maximum ephemeris difference over the integrated periods. The chosen summary statistics give the user a sense of the typical differences, as well as the maximum differences in predicted positions over the integration period.

\subsection{Test Objects} \label{subsec:objects}
We have selected 22 objects to represent different populations for orbit propagation. Our set includes Main Belt Asteroids (MBAs), Potentially Hazardous Asteroids (PHAs), Near Earth Objects (NEOs), Trans-Neptunian Objects (TNOs), and representatives of Mars Trojans, Jupiter Trojans, Centaurs, Impactors, a ``Vatira'' asteroid (2020 AV2), and an Interstellar Object (`Oumuamua). These were chosen with LSST applications in mind, but \texttt{iCompare} allows the user to modify the list as necessary.

\subsection{Binder support}

We provide the ability to directly launch and test a Jupyter notebook with \texttt{iCompare} using mybinder.org, an ``online service for building and sharing reproducible and interactive computational environments from online repositories''\footnote{\url{https://mybinder.readthedocs.io/en/latest/introduction.html}}. This makes it possible to try and run \texttt{iCompare} without installing any software locally.

\section{Results} \label{sec:result}



\begin{figure*}
    \centering
    \includegraphics[scale=0.75]{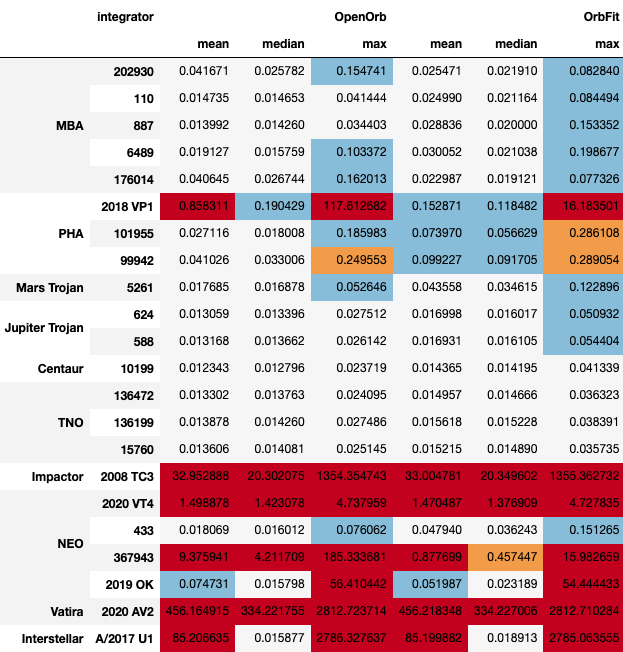}
    \caption{A screenshot of an \texttt{iCompare}-generated table of differences between ephemerides (in arcseconds), color-coded with the following schema: red $>$ 1''; orange $>$ 0.6''; blue $>$ 0.2''; transparent $>$ 0.05''}
    \label{fig: table1}
\end{figure*}

Figure~\ref{fig: table1} shows an example run of \texttt{iCompare}, using the default settings and test objects as described in preceding sections. The columns contain integrator and statistical value labels, and the rows show the asteroids, grouped by object type. All values in the table are expressed in arcseconds.
The data is color-coded as follows: if the difference between ephemerides is less than 0.05 arcseconds, no color is applied. If the difference up to 0.2" (roughly a single LSSTCam pixel), the cell is blue to denote a small but statistically detectable difference. A difference up to 0.6" (comparable to Rubin Observatory median PSF width of 0.8") is marked orange. Finally, the cells shaded red indicate significant differences between integrators that are likely to be observable even in single-epoch observations. As can be inferred from the previous paragraphs, the default color map has been selected for Rubin Observatory development purposes. The user can override it via an argument to the \lstinline{icompare.plot()} function.

Much of the table has no color flags, showing close agreement of all tested integrators. As expected, the rows towards the bottom -- impactors, NEOs, and an ISO with cometary behavior -- show the largest deviations. These are particularly difficult objects to integrate, especially with default settings. While individual object accuracy could be improved on via hand-tuning the settings, we intentionally avoid doing so since the LSST is meant to be an automated survey. Instead, we hope to encourage the integrator writers to add functionality to auto-select the appropriate defaults based on the object being integrated.


\section{Conclusions and Future Work} \label{sec:conc}

This note introduces \texttt{iCompare}, a Python package for the comparison of the most popular integrators in Solar System dynamics to JPL/HORIZONS. 
It enables a rapid visual comparison of ephemerides prediction accuracy of Solar System integrators across a range of orbits and object types. It is modular and extensible to new integrators by additions of a single Python wrapper. We hope \texttt{iCompare} will be useful to both integration package authors and users selecting an integrator for their specific application. We would especially encourage improvements towards automation of the integration packages such that they require less per-object hand-tuning; this is crucial for applications in automated surveys such as LSST.

\texttt{iCompare} is available at \url{https://github.com/dirac-institute/iCompare}; feedback, feature suggestion, and pull requests are welcome and we invite collaboration on future improvements.

\software{astropy \citep{Astropy2013},
          astroquery \citep{astroquery},
          pandas \citep{pandas-soft},
          openorb/pyoorb \citep{openorb},
          orbfit (\url{http://adams.dm.unipi.it/orbfit/}),
          jpl/horizons (\url{https://ssd.jpl.nasa.gov/?horizons}) 
          }


\bibliography{iCompareNote}{}
\bibliographystyle{aasjournal}



\end{document}